\documentclass[journal]{IEEEtran}
\usepackage{amssymb}
\usepackage{graphicx}
\usepackage{subcaption}
\graphicspath{{images/}}
\usepackage{multirow}
\usepackage{url}    

\begin{document}

\title{On Filter Generalization for Music Bandwidth Extension Using Deep Neural Networks}

\author{
    Serkan Sulun and Matthew E.~P.~Davies
    \thanks{Serkan Sulun is with the Institute for Systems and Computer Engineering, Technology and Science (INESC TEC), 4200–465 Porto, Portugal (e-mail: serkan.sulun@inesctec.pt).
    Matthew E. P. Davies is with the University of Coimbra, Centre for Informatics and Systems of the University of Coimbra, Department of Informatics Engineering, Portugal
    (email: mepdavies@dei.uc.pt).
    Serkan Sulun receives the support of a fellowship from ”la Caixa” Foundation (ID 100010434), with the fellowship code LCF/BQ/DI19/11730032.
    This work is funded by national funds through the FCT - Foundation for Science and Technology, I.P., within the scope of the project CISUC - UID/CEC/00326/2020 and by European Social Fund, through the Regional Operational Program Centro 2020 as well as by Portuguese National Funds through the FCT - Foundation for Science and Technology, I.P., under the project IF/01566/2015.}
    }
\maketitle

\begin{abstract}
In this paper, we address a sub-topic of the broad domain of audio enhancement, namely musical audio bandwidth extension. We formulate the bandwidth extension problem using deep neural networks, where a band-limited signal is provided as input to the network, with the goal of reconstructing a full-bandwidth output. Our main contribution centers on the impact of the choice of low pass filter when training and subsequently testing the network. For two different state of the art deep architectures, ResNet and U-Net, we demonstrate that when the training and testing filters are matched, improvements in signal-to-noise ratio (SNR) of up to $7$\,dB can be obtained. However, when these filters differ, the improvement falls considerably and under some training conditions results in a lower SNR than the band-limited input. To circumvent this apparent overfitting to filter shape, we propose a data augmentation strategy which utilizes multiple low pass filters during training and leads to improved generalization to unseen filtering conditions at test time.  
\end{abstract}


\section{Introduction}
\label{S:1}

\IEEEPARstart{M}{odern} recording techniques provide music signals with extremely high audio quality. By contrast, the listening experience of archive recordings, such as jazz, pop, folk, and blues recorded before the 1960s is arguably limited by the recording techniques of the time as well as the degradation of physical media. Even so, modern recordings can also suffer from diminished audio quality due to the use of lossy compression, downsampling, packet loss, or clipping. In the broadest sense, audio enhancement aims to restore a degraded signal to improve its sound quality \cite{restoration}. As such, audio enhancement may target the removal of noise, the suppression of cracks or pops (e.g. from old vinyl records), signal completion to fill in gaps (so-called ``audio inpainting'' \cite{audioinpainting,perraudin2018inpainting}), or the bandwidth extension of a band-limited signal.

To transmit audio signals through internet streams, or for the ease of storing, common operations such as compression, bandwidth reduction, and low-pass filtering all result in the removal of at least part of the high-frequency audio content. Optionally, the signal can be downsampled afterwards, effectively reducing its size. While this process can be understood as a relatively straightforward mapping from a \textit{full-bandwidth}, or \textit{wideband} signal to a \textit{band-limited} or \textit{narrowband} signal, the corresponding inverse problem, namely \textit{bandwidth extension}, seeks to reconstruct missing high-frequency content and is thus non-trivial. Furthermore, if the input signal is downsampled, the inverse problem also requires upsampling, and the overall process is called \textit{super-resolution}, a term that is commonly used in the image processing literature. Despite these challenges, bandwidth extension is crucial for increasing the fidelity of audio, especially for speech and music signals.

The first applications of audio bandwidth extension addressed speech signals only, due to the practical problems arising from the low bandwidth of telephone systems. One of the earliest works used a statistical approach in which narrowband and wideband spectral envelopes were assumed to be generated by a mixture of narrowband and wideband sources \cite{cheng1994}. Codebook mapping-based methods use two learned codebooks, belonging to the narrowband and wideband signals, containing spectral envelope features, where a one-to-one mapping exists between their entries \cite{yoshida1994,epps1999}. In linear mapping-based methods, a transformation matrix is learned using methods such as least-squares \cite{nakatoh1997,chennoukh2001}.

Later methods sought to learn to model the wideband signal directly, rather than the mapping between predefined features. Gaussian mixture models (GMMs) have been used to estimate the joint probability density of narrowband and wideband signals \cite{park2000,nour2008}. Other approaches include the use of hidden Markov models (HMMs), where each state of the model represents the wideband extension of its narrowband input \cite{jax2003,bauer2008,song2009}. Due to its recursive mechanism, HMMs can leverage information from the past input frames. Methods based on non-negative matrix factorization (NMF) model the speech signals as a combination of learned non-negative bases \cite{bansal2005,sun2013}. In the testing stage, low-frequency base components of the input can be used to estimate how to combine the high-frequency base components to create the wideband signal. Finally, the first works using neural networks for speech bandwidth extension employed multilayer perceptrons (MLPs) to estimate linear predictive coding (LPC) coefficients of the wideband speech signal \cite{iser2003}, or to find a shaping function to transform the spectral magnitude \cite{kontio2007}. We note that these early works used very small neural networks, in which the total number of parameters was around $100$.

More recent approaches to audio bandwidth extension have used deep neural networks (DNNs), with many more layers and far greater representation power than their older counterparts. DNNs also eliminate the need for hand-crafted features, as they can use raw audio or time-frequency transforms as input, and then learn appropriate intermediate representations. Early works using DNNs on speech bandwidth extension employed audio features as inputs, and demonstrated the superiority of DNNs over the state-of-the-art method of the time, namely GMMs \cite{against_gmm, against_gmm2}. Another pioneering DNN-based work used the frequency spectrogram as the input \cite{li}. A much deeper model employed the popular \textit{U-Net} architecture \cite{unet} and works in the raw audio domain, performing experiments on both speech and single instrument music \cite{kuleshov}. Lim et al. combined the two aforementioned approaches creating a dual network, which operates separately in the time and frequency domains, and creates the final output using a fusion layer \cite{timefrequency}. A recent work used the \textit{U-Net} in the time domain only, but the training loss was a combination of losses calculated in both the time and frequency domains \cite{tfloss}. To increase the qualitative performance, namely, the clarity of the produced audio, generative adversarial networks \cite{gan} have also been employed in DNN-based audio bandwidth enhancement \cite{bwe_gan, bwe_gan2}. The latest work by Google on music enhancement presents an ablation study, using SNR to measure distortion, and VGG distance, namely the distance between the embeddings of the VGGish network \cite{vggish}, as the perceptual score \cite{denoising}. Their results show that the incorporation of adversarial loss yields a better perceptual score at the expense of decreasing SNR.

\section{Motivation and Paper Outline}

\subsection{Motivation}

While the enhancement of old music recordings can be partially framed in the context of bandwidth extension, certain risks arise when considering the data that DNNs are given for training. Even though trained DNNs can perform well on samples from the training data, they may not exhibit the same performance on unseen samples from the testing data. This phenomenon is named \textit{sample overfitting} and even though it is an important concern, especially for classification tasks, its existence in generative tasks, such as image super-resolution, audio bandwidth extension, and adversarial generation, is debated. Recent studies show that sample overfitting is not observed for both discriminators and generators of generative adversarial networks \cite{gan_overfitting1, gan_overfitting2}, and supervised generative networks for video frame generation \cite{serkan}. Furthermore, state-of-the-art image super-resolution networks do not include any regularization layers \cite{imagesr1,edsr,imagesr3}, such as batch normalization \cite{batchnorm} and dropout \cite{dropout}, to avoid overfitting.

Especially in the task of automatic speech recognition, models may not generalize well to audio samples recorded in a completely different environment, even when the speakers remain the same. Methods to resolve this problem are referred in the literature as \textit{multi-environment} \cite{multi_environment}, \textit{multi-domain} \cite{multi_domain}, or \textit{multi-condition} \cite{multi_condition} approaches, and consist of using training samples recorded in multiple environments, with the goal of generalization to unseen environments. Some works simulate the multiple environment conditions through data pre-processing. One study created training samples by adding noise with different signal-to-noise (SNR) levels on clean speech signals \cite{multi_snr}. Another work on speech bandwidth enhancement included input training samples that are created using low-pass filters with different cut-off frequencies \cite{multi_cutoff}. In all aforementioned examples, the samples that illustrate multiple conditions are perceptually different.

Another risk concerns the pre-processing methods used to create the training data. When considering music bandwidth extension for enhancing archive recordings, no full-bandwidth version exists and as such, there is no ``ground-truth'' target for DNNs. To this end, training data is typically obtained by low-pass filtering full-bandwidth recordings. However, since real-world band-limited samples are not the result of some hypothetical universal digital low-pass filter, it can be challenging to develop robust techniques for bandwidth extension which rely on a loose approximation of the bandwidth reduction process, and in turn to generalize to unseen recordings. While trained DNNs perform well on training data created with one type of low-pass filter, they may fail to generalize to audio content subjected to different types of low-pass filters. This phenomenon can occur even when these different types of low-pass filters have the same cut-off frequency, creating samples that may have almost no perceptual difference. Throughout this paper, we call this \textit{filter overfitting}, which can be understood as a lack of \textit{filter generalization}.

While Kuleshov et al. \cite{kuleshov} do not explicitly target filter generalization, they present a rudimentary analysis of generalization related to the presence or absence of a pre-processing filter. Their main goal is audio super-resolution, and while preparing their band-limited input data, before downsampling, they optionally use a low-pass filter. They demonstrate results in which a low-pass filter is not present while preparing the input training data, but is present for the test data, and vice-versa. Both training and testing data are still downsampled, hence they investigate the generalization in the context of aliased and non-aliased data. When the aliasing conditions match, the model performs well, with test SNR levels around $30$\,dB. But when these conditions do not match, the model becomes ineffective, with test SNR levels around $0.4$\,dB, showing no generalization to the addition or the removal of the low-pass filter during testing.

\subsection{Contributions}

While the use of low pass filtering is widespread among existing work on audio bandwidth extension using DNNs, to the best of our knowledge, no work to date has thoroughly investigated the topic of filter generalization. We argue that the lack of generalization to various types of signal deterioration is an important challenge in creating audio enhancement models for real-world deployment. In this work, we present a rigorous analysis of filter generalization, evaluating generalization to different filters used to pre-process input data, on the task of bandwidth enhancement of complex music signals, using two popular DNN architectures.

To evaluate sample overfitting, we disjoint testing and training data, to create totally \textit{unseen data} for the trained models. To evaluate filter generalization, we pre-process the testing input data with a filter that does not match the filters that pre-process the training input data, i.e., an \textit{unseen filter} and compare it to the test setting where the filters used for training and testing data do match, i.e, \textit{seen filters}. We argue that testing with the unseen filter can be considered a kind of real-world signal degradation, in which the true underlying degradation function is unknown.

We evaluate three different regularization methods that are used in the literature to increase generalization. In particular, we compare the usage of data augmentation, batch normalization, and dropout, against the baseline of not using any regularization methods. We introduce a novel data augmentation technique of using a set of different low pass filters to pre-process the input data, in which the unseen test filter is never present. We examine the training process by tracking the model's performance throughout training iterations, by performing validation using both seen and unseen filters.

Similar to image super-resolution methods, we use fully-convolutional DNNs to model the raw signal directly \cite{edsr, image_sr}. One of the DNN models we employ is the \textit{U-Net}, which was first used for biomedical image segmentation \cite{unet}, and later in audio signal processing tasks such as singing voice separation \cite{audiounet}, and eventually for audio enhancement \cite{kuleshov, bwe_gan, tfnet, waveunet}. In addition to the U-Net, we also use the deep residual network model (\textit{ResNet}) \cite{resnet} since it is one of the most widely used DNN architectures in signal processing tasks. Even though the U-Net is a popular architecture in the recent audio processing literature, to the best of our knowledge, no work in the domain of audio processing compares the U-Net against the well-established baseline of the ResNet. A small number of comparative studies exist in the fields of image processing and medical imaging, in which either the number of parameters of the compared models is not stated \cite{celltracing}, or in which the ResNet has significantly fewer parameters than the U-Net \cite{mr,hdr,cancer}. In all these works, the ResNet outperforms the U-Net by a small margin. In this paper, we also present a comparison between the U-Net and ResNet, where each network has a similar number of parameters.

Our main findings indicate that filter overfitting occurs for both the U-Net and ResNet, although to different degrees, and that the use of multi-filter data augmentation, as opposed to more traditional regularization techniques, is a promising means to mitigate this overfitting problem and thus improve filter generalization for bandwidth extension.

\subsection{Outline}

The remainder of the paper is organized as follows. Section \ref{sec:models} presents the architectures of the baseline models used. Section \ref{sec:regularization} defines the existing regularization layers for DNNs and introduces our novel data augmentation method. The rest of Section \ref{sec:methodology} describes the dataset, evaluation methods, and implementation details. In Section \ref{sec:results} we present a detailed analysis of the performance of the trained models. Finally, in Section \ref{sec:conclusions} we present conclusions and highlight promising areas for future work.

\section{Methodology}
\label{sec:methodology}

\subsection{Models}
\label{sec:models}

In this section, we define the two baseline models: U-Net and ResNet. For both models, we follow the approach of Kuleshov et al. \cite{kuleshov} and use raw audio as the input rather than time-frequency transforms (e.g., as in \cite{matthew}). As such we remove any need for phase reconstruction in the output. However, since we address bandwidth extension and not audio super-resolution, our inputs are not subsampled. Hence the sizes of the input and the output are equal for all our models.

\subsubsection{U-Net}
\label{sec:unet}

\begin{figure}[ht]
\centering
\begin{subfigure}{.24\textwidth}
  \centering
 
\includegraphics[width=.9\textwidth]{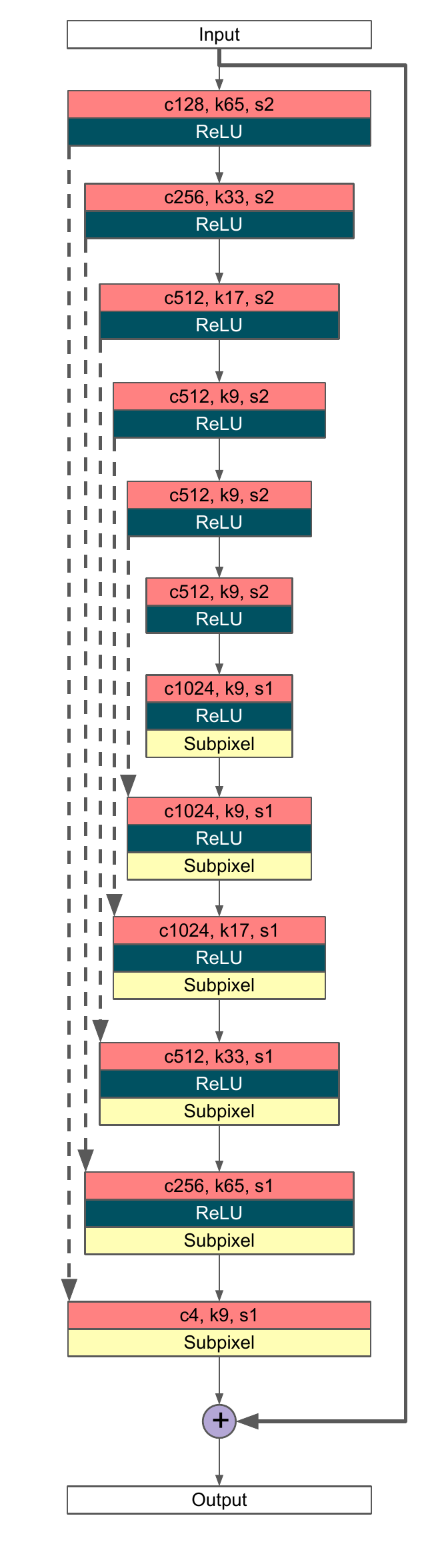}  
  \caption{U-Net model. Dashed lines \\ indicate stacking connections.}
  \label{fig:unet}
\end{subfigure}
\begin{subfigure}{.24\textwidth}
  \centering
 
\includegraphics[width=.9\textwidth]{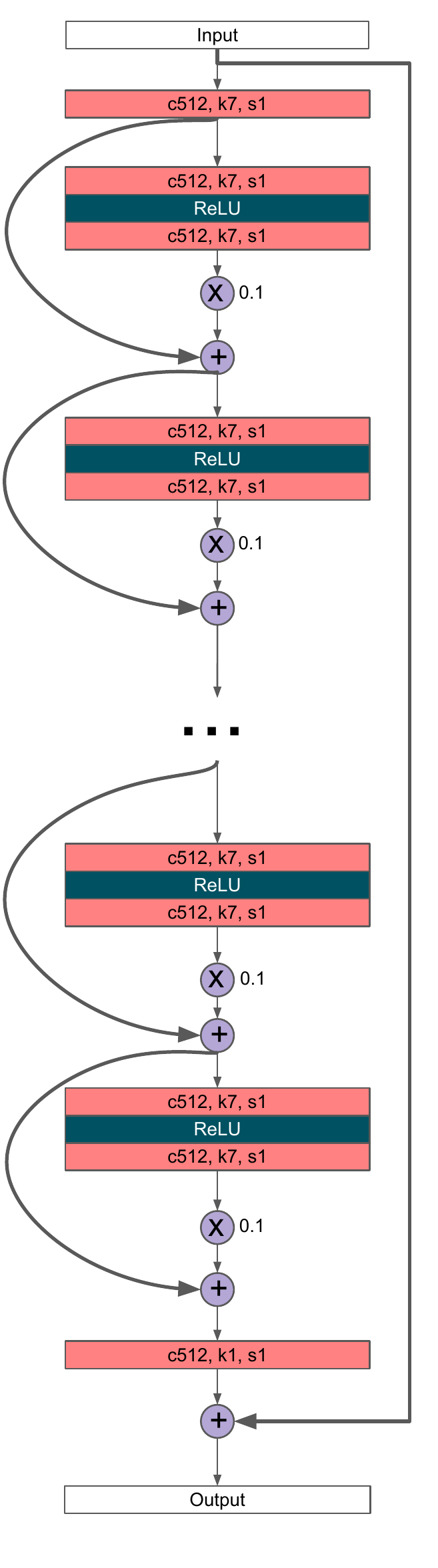}  
  \caption{ResNet model with 15 residual blocks.}
  \label{fig:resnet}
\end{subfigure}
\caption{Models used. c, k, and s indicate channel size, kernel size and stride of the convolutional layers, respectively.}
\label{fig:models}
\end{figure}

The U-Net architecture \cite{unet}, like the auto-encoder, consists of two main groups. The first group contains downsampling layers and is followed in the second group by upsampling layers, as shown in Figure \ref{fig:unet}. In the U-Net, individual downsampling and upsampling layers at the same scale are connected through stacking connections, e.g., the output of the first downsampling convolutional block is stacked with the input of the last upsampling convolutional block.

In the downsampling group, one-dimensional convolutional layers with stride $2$ are used, effectively halving the activation length. Borrowing from image processing terminology, the upsampling group includes ``sub-pixel'' layers (also known as the pixel shuffler) \cite{subpixel} to double the activation length. Sub-pixel layers weave the samples in the spatial dimension, taken from alternate channels, effectively halving the channel length.

The number of parameters is selected to replicate the original work using U-Net for audio super-resolution \cite{kuleshov, github}, which we denote as \textit{Audio-SR-U-Net} throughout this paper. This resulted in a network with $56.4$ million parameters.

\subsubsection{ResNet}
\label{sec:resnet}

A common issue with training vanilla feed-forward neural networks with many layers is the ``vanishing gradient" problem, in which the gradient back-propagated to the earliest layers approaches zero, due to repeated multiplications. Residual networks \cite{resnet} eliminated this problem by using \textit{residual blocks}, which only model a fraction of the difference between their inputs and outputs. Commonly, each residual block includes two convolutional layers and a nonlinear function in between them. Very deep models include \textit{residual scaling} in which the output of each residual block is multiplied by a small number, e.g., $0.1$, and then summed with its input, to further stabilize training. Our ResNet model is represented in Figure \ref{fig:resnet}.

Unlike the U-Net, the ResNet activation lengths stay constant throughout the network. In this way, we can avoid any loss of temporal information since our goal is to create a high-resolution output of equal length to the input. Note that we use a simple design where all convolutional layers except the last one have the same number of parameters. Similar in size to the U-Net implementation, it has $55.1$ million parameters.

In all our models, all convolutions apply appropriate zero padding to keep the activation sizes constant. This is even true for the downsampling convolutions since the downsampling effect is achieved using strided convolutions. The \textit{Rectified Linear Unit (ReLU)} is used as the activation function. The loss function for all our models is the mean-squared error. As is common in enhancement models, an additive connection from the input to the output is also used, so that the network only needs to model the \textit{difference} between the input and the target signals, rather than creating the target signal from scratch.

To analyze generalization, we present ablation studies, in which we incorporate common methods to avoid overfitting, defined as \textit{regularization methods}.

\subsection{Regularization methods}
\label{sec:regularization}

\subsubsection{Dropout}

One of the simplest methods to prevent overfitting is dropout, where activation units are dropped based on a fixed probability \cite{dropout}. This introduces noise in the hidden layers and prevents excessive co-adaptation.

Although dropout has been largely superseded by batch normalization, especially in residual networks, new state-of-the-art residual models, namely wide residual networks \cite{wide_resnet} do employ it. Furthermore, \textit{Audio-SR-U-Net}'s open-source implementation \cite{github} uses a dropout layer instead of batch normalization, and thus, we followed this approach in our U-Net model and used dropout layers after each upsampling convolutional layer. In our ResNet model, we placed dropout layers between the two convolutional layers of each residual block. For all experiments, the dropout rate is set to $0.5$.

\subsubsection{Batch normalization}

While training DNNs, updating the parameters of the model effectively changes the distribution of the inputs for the next layers. This is defined as \textit{internal covariate shift} and batch normalization addresses this problem by normalizing the layer inputs \cite{batchnorm}. Even though batch normalization is mainly proposed to speed up training, it provides regularization as well. Because the parameters for the normalization are learned based on each batch, they can only provide a noisy estimate of the true mean and variance. Normalization using these estimated parameters introduces noise within the hidden layers and reduces overfitting.

For the U-Net, we follow the \textit{Audio-SR-U-Net} model \cite{kuleshov} and insert batch normalization layers after each downsampling convolutional layer. For the ResNet, batch normalization is used after each convolutional layer.

\subsubsection{Data augmentation}

To increase sample generalization of DNNs, data augmentation is used, where the input data samples are transformed before being fed into the DNN, effectively increasing the number and diversity of training samples. Data augmentation is very common in image-based tasks and mostly utilizes geometric transformations such as rotating, flipping, or cropping \cite{dataaugmentation}. Geometric transformations of this kind when applied to music signals typically do not produce realistic samples. While some work has been conducted on data augmentation for musical signals \cite{mcfee2015_augmentation}, it primarily targets robustness for classification tasks such as instrument identification in the presence of time-stretching, pitch-shifting, dynamic range compression, and additive noise. Operations of this kind (including minor changes in time or pitch) certainly form part of a larger set of signal degradations that could be explored for musical audio enhancement, however, our focus in this work centers on bandwidth extension and is thus restricted to the consideration of low pass filtering.

\begin{table}[t!]
\centering
\begin{tabular}{c|c|c}
                               
& \begin{tabular}[c]{@{}c@{}}Single-filter\\ (No data augmentation) \vspace{1mm}\end{tabular}  & \begin{tabular}[c]{@{}c@{}}Multi-filter\\ (Data augmentation) \vspace{1mm}\end{tabular}                                                                                   
\\ \hline
\begin{tabular}[c]{@{}c@{}} \vspace{1mm} \\ Training \vspace{4mm}  \\ \hline \\ \vspace{-2mm} \\ Validation with\\ seen filter(s) \\ \vspace{-2mm} \end{tabular}                & Chebyshev-$1$, $6$  & \begin{tabular}[c]{@{}c@{}} \vspace{-2mm} \\ Chebyshev-$1$, $6$\\ Chebyshev-$1$, $8$\\ Chebyshev-$1$, $10$\\ Chebyshev-$1$, $12$\\ Bessel, $6$\\ Bessel, $12$\\ Elliptic, $6$\\ Elliptic, $12$ \vspace{1mm}\end{tabular} \\ \hline
\begin{tabular}[c]{@{}c@{}} \vspace{-2mm} \\ Validation with\\ unseen filter \\ \vspace{-2mm} \\ \hline \\ \vspace{-5mm} \\ Testing with\\ unseen filter \\ \vspace{-2mm} \end{tabular} & Butterworth, $6$ & Butterworth, $6$                                                                                                                                   
\\ \hline
\begin{tabular}[c]{@{}c@{}}\vspace{-2mm} \\ Testing with\\ seen filter \\ \vspace{-2mm}\end{tabular}                                        & Chebyshev-$1$, $6$  & Chebyshev-$1$, $6$                                                      
\end{tabular}

\caption{The types and orders of the low-pass filters used, under two different training settings, \textit{single-filter} (no data augmentation) and \textit{multi-filter} (data augmentation).}
\label{table:filters}
\end{table}

\begin{figure}[!ht]
\begin{center}
  \includegraphics[width=\linewidth]{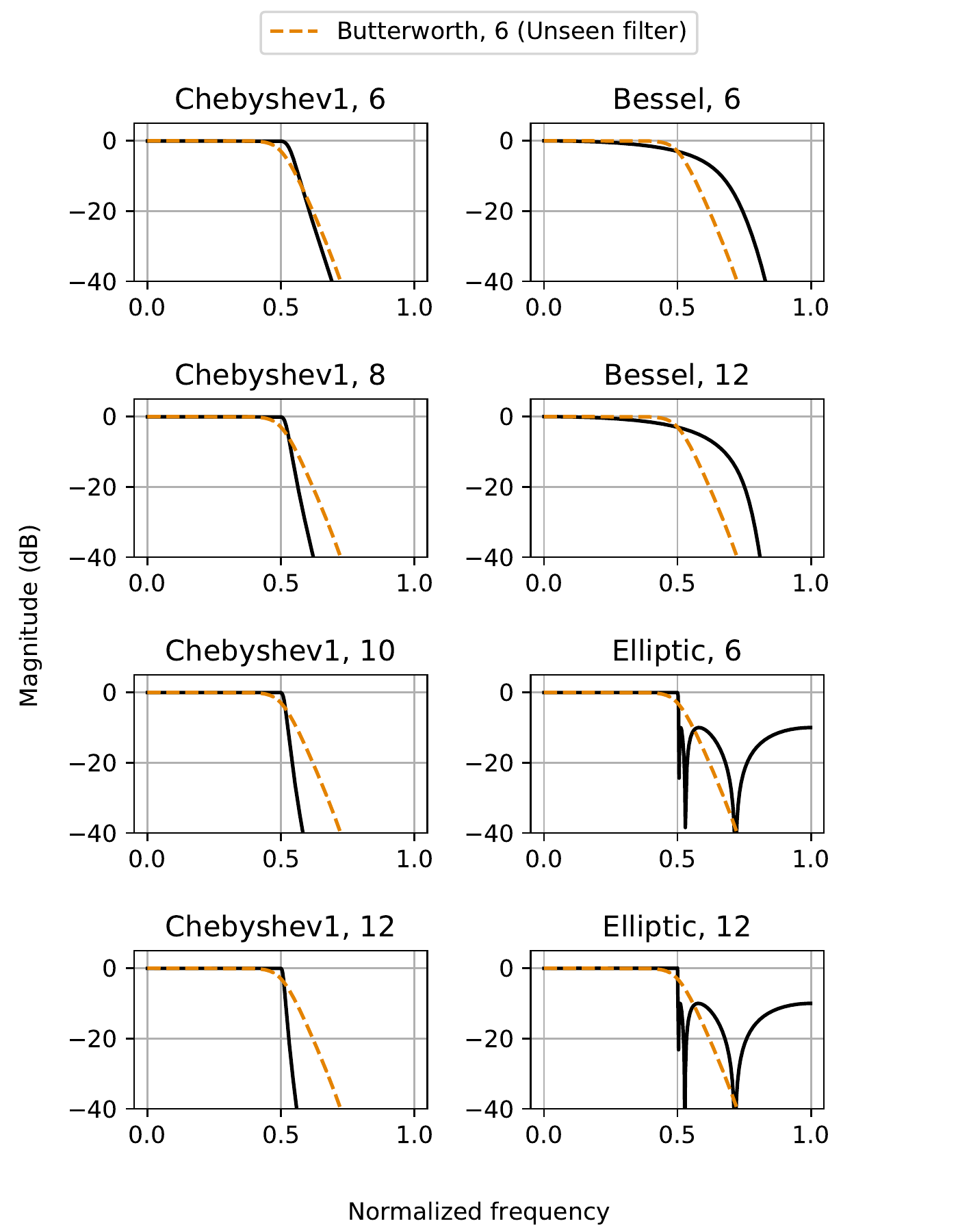}
\end{center}
\caption{Frequency responses of the training filters. The frequency response of the unseen filter, $6$th order Butterworth is superimposed on each plot.}
\label{fig:filters}
\end{figure}

Since our main goal in this work is to explore and then improve filter generalization, we propose a data augmentation method where many different types of filters are used during training. Our baseline approach, without data augmentation, uses a \textit{single-filter} training setting, specifically a $6$th order Chebyshev Type $1$, denoted as ``Chebyshev-$1$, $6$". When using data augmentation, in a \textit{multi-filter} training setting, we adopt a set of eight different filters, picked randomly for each input sample during training. These eight filters consist of \textit{Chebyshev-$1$}, \textit{Bessel}, and \textit{Elliptic} filters of different orders. To evaluate filter generalization, we reserve the $6$th order \textit{Butterworth} filter as the unseen filter. The filters are summarized in Table \ref{table:filters}, and their usage during evaluation is detailed in Section \ref{sec:eval}. A graphical overview of their different frequency magnitude responses is shown in Figure~\ref{fig:filters}.

\subsection{Dataset}

Machine learning approaches to bandwidth extension formulate the problem via the use of datasets that contain both full-bandwidth (high-quality) and band-limited (low-quality) versions of each audio signal. A straightforward way to construct these pairs of samples is to obtain a high-quality dataset and then to low-pass filter it. Even though there are many musical audio datasets, especially within the music information retrieval community, many of them are collated from diverse sources (including researchers' personal audio collections) and often contain audio content has been compressed (e.g. via lossy MP3/AAC encoding), hence they are not strictly full-bandwidth nor easily reproducible.

Other than the need for full-bandwidth musical audio content, our proposed approach is intended to be agnostic to musical style. To this end, any uncompressed full-bandwidth musical content could be used as training material, however, to allow  reproducibility, we select the following two publicly available datasets, which contain full-bandwidth, stereo, and multi-track musical audio: MedleyDB (version 2.0) \cite{medleydb} and DSD100 \cite{dsd100}. In each dataset, the audio content is sampled at $44100$\,Hz, with a bandwidth of $22050$\,Hz.

MedleyDB consists of $121$ songs, while DSD100 has two splits for training and testing, each containing $50$ songs. Given the inclusion of isolated multi-track stems, both datasets have found high uptake in music mixing and audio source separation research. However, in this work, we seek to address bandwidth extension for multi-instrument music as opposed to isolated single instruments, and thus we retain only the stereo mixes of each song. To create band-limited input samples, we apply a low pass filter with a fixed cut-off of $11025$\,Hz, i.e., half the bandwidth of the original. Dataset samples contain values ranging from $-1$ to $1$ and we haven't performed any additional pre-processing, e.g., loudness normalization.

The DSD100 test split is used for testing, the last $8$ songs of DSD100 training split are used for validation, with all remaining songs of DSD100 training split plus the entire MedleyDB dataset are used for training. On this basis, the training, validation, and testing sets are all disjoint.

\subsection{Evaluation}
\label{sec:eval}

\subsubsection{Metrics}

To measure the overall distortion of the outputs, we use the well-established  signal-to-noise ratio (SNR):

\begin{equation}
\textrm{SNR}(x,\hat{x}) = 10 \log_{10} \frac{||x||_2^2}{||x-\hat{x}||_2^2}
\end{equation}

where $x$ is the reference signal and $\hat{x}$ is its approximation. While calculating the $2$-norms, the signals are used in their stereo forms. In the specific context of our work, we consider SNR to be an appropriate choice to investigate overfitting since our models are trained with the mean-squared loss, and minimizing it corresponds to maximizing SNR.

To provide additional insight into performance, we evaluate the perceptual quality of the output audio samples, using the VGG distance, as used recently by Li et al. for the evaluation of music enhancement \cite{denoising}. The VGG distance between two audio samples is defined as the distance between their embeddings created by the \textit{VGGish} network pre-trained on audio classification \cite{vggish}. 
A recent work on speech processing shows that the distance between deep embeddings correlates better to human evaluation, compared to hand-crafted metrics such as Perceptual Evaluation of Speech Quality (PESQ) \cite{pesq} and the Virtual Speech Quality Objective Listener (ViSQOL) \cite{visqol}, across various audio enhancement tasks including bandwidth extension \cite{perceptual_metric}. The \textit{VGGish} embeddings are also used in measuring the Fréchet Audio Distance (FAD), a state-of-the-art evaluation method to assess the perceptual quality of a collection of output samples \cite{fad}. However, because FAD is used to compare two collections rather than individual audio signals, it is not applicable in our case.

To obtain the VGG embeddings, we used the \textit{VGGish} network's open-source implementation \cite{vgg_github}. We used the default parameters except setting the sampling frequency to $44100$\,Hz and the maximum frequency to $22050$\,Hz. In contrast to the SNR calculation, the reference implementation downmixes the stereo signals to mono before calculating the VGG embeddings. After post-processing, the embeddings take values from $0$ to $255$.  Similar to Manocha et al. \cite{perceptual_metric}, we employ the mean absolute distance to define the VGG distance as:

\begin{equation}
\textrm{VGG}(x,\hat{x}) = \frac{1}{n} \sum_{i=1}^n |y_i - \hat{y}_i |
\end{equation}

where $x$ is the reference signal, $\hat{x}$ is its approximation; $y$ and $\hat{y}$ are their embeddings, respectively. $n$ is the size of the embedding tensors, which depends on the length of $x$.

Given the need to make a large number of objective measurements throughout the training and testing (as detailed in Section \ref{sec:results}), we do not pursue any subjective listening experiment and leave this as a topic for future work.  

\subsubsection{Testing}

To assess the overall performance of our models, we perform testing once, at the end of the training. The test split of the DSD100 dataset is reserved for our testing stage. Due to GPU memory limitations, our networks cannot process full-length songs in a single forward pass, hence they process non-overlapping chunks of audio and the outputs are later concatenated to create full-length output songs. For both VGG distance and SNR, we calculate them at the song level first, based on these full-length songs, and then take the mean over the data split to obtain the final test values.

To evaluate filter generalization, we perform two tests for each model, using seen and unseen filters. As summarized in Table \ref{table:filters}, the $6$th order Butterworth filter is selected as the unseen filter, as it is not used in any training setting. The seen filter only includes $6$th order Chebyshev-$1$, as this is the only filter common to both single and multi-filter training settings.

\subsubsection{Validation}
\label{sssec:validation}

To observe generalization or overfitting throughout training, we perform validation repeatedly, where we measure the output SNR once every $2500$ training iterations. We perform validation on $8$\,s audio excerpts, starting from the $8$th second of each song, for only $8$ songs. These $8$ songs correspond to the last $8$ of the DSD100 training split. Since the validation is performed repeatedly throughout training, we keep the validation set sample size small. We believe that this small sample size is sufficient, because validation is only used to observe the progress of training, and the final performance evaluation is done in the testing stage. The final validation SNR is obtained by first calculating it over each $8$\,s, and then taking the mean over the validation songs.

Similar to testing, the validation is also performed twice, using seen and unseen filters. Validation with the unseen filter uses the $6$th order Butterworth filter, as in testing. Because validation with the seen filter(s) is done to observe the training progress of each model and not to compare different models, the filters employed are the same as those in the corresponding training setting. As seen in Table \ref{table:filters}, in the single-filter setting, validation with the seen filter only has the Chebyshev-$1$ filter, and in the multi-filter setting, it uses all eight training filters, with each assigned to processing a different song in the validation data split.

\subsection{Implementation details}

We built and trained our models using the Pytorch framework \cite{pytorch} and a single Nvidia GeForce GTX 1080 Ti GPU. The model weights are initialized randomly with values drawn from the normal distribution with zero mean and unit variance. The batch size is $8$. We use the Adam optimizer \cite{adam} with an initial learning rate of $5$e-$4$, and with beta values 0.9 and 0.999. The learning rate is halved when the training loss reaches a plateau. We record the average training loss every $2500$ iterations, and consider a plateau to correspond to no decrease in loss for $5$ such consecutive measurements. Training samples are created by first randomly picking an audio file from the training dataset and then, at a random location in the audio file, extracting a chunk of stereo audio, with a length of $8192$ samples, corresponding to $186$ milliseconds. However, since all our models are fully-convolutional, they can process audio signals with arbitrary lengths. We train our models until convergence and for testing we use the model weights taken from the conclusion of the training. Our source code is available online\footnote{\url{https://github.com/serkansulun/deep-music-enhancer}}.

\section{Results}
\label{sec:results}

\begin{figure*}[h!]
\begin{center}
 
\includegraphics[width=0.9\textwidth]{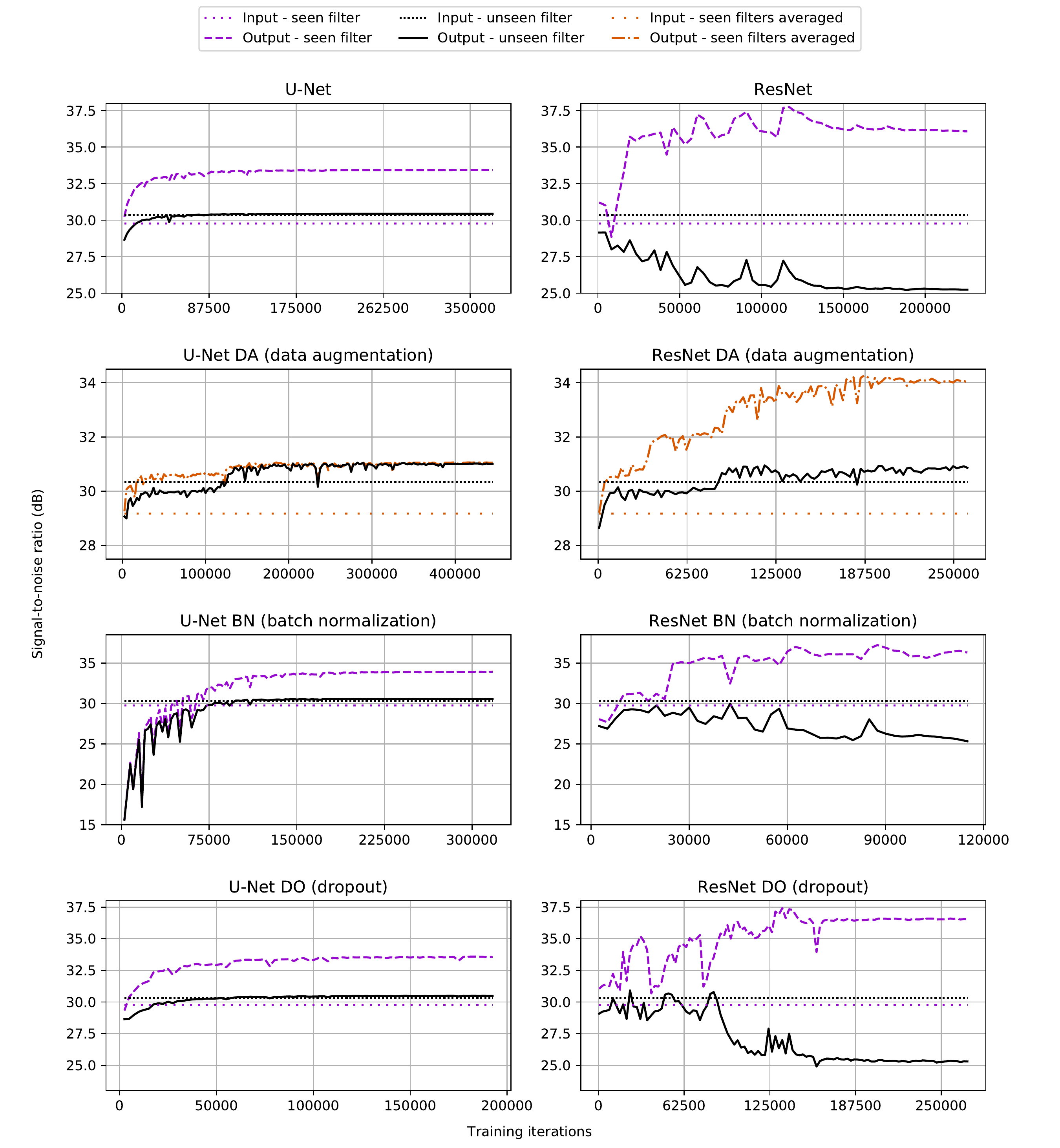}
\end{center}
\caption{Validation performance of our models throughout training. The input and output SNR levels are measured by comparing input and model output samples against the ground-truth. Since the inputs are not affected by training, their SNR level stays constant throughout and constitutes a baseline. The seen and unseen filters are detailed in Table \ref{table:filters}. For the data augmentation experiments, there are multiple seen filters, and the SNR levels are computed by taking the average across multiple filters.}
\label{fig:all}
\end{figure*}

\subsection{Validation Data}

Figure \ref{fig:all} provides a high-level overview of the performance of all of the different models and training schemes, with the SNR as a function of the training iterations. While the horizontal dashed lines indicate a baseline of input SNR levels, the rest of the lines denote output SNRs for both validation settings. The SNR levels of the input validation with seen filter(s) are different for the experiments with data augmentation since a different number of training filters are used as summarized in Table \ref{table:filters}, and as shown in Figure \ref{fig:filters} their differing frequency responses naturally lead to different baseline SNRs.

Examining the first row of Figure \ref{fig:all} we see that for both networks, when the input filter is known, then large improvements in SNR over the baseline are possible. However, contrasting the U-Net with the ResNet, the performance with the unseen filter is markedly different. For the U-Net the output SNR converges to the baseline, but for the ResNet, performance degrades as training continues. In this way, we see quite clear evidence of a lack of filter generalization in both models.

Moving to the second row, where training includes data augmentation, we observe a different pattern, where both networks improve upon the baseline SNR for the unseen filter. Contrasting the U-Net and ResNet, we see that the ResNet offers a greater improvement upon the set of seen filters than the U-Net, albeit for approximately the same number of parameters.

Inspection of the third and fourth rows which include the two regularization techniques, we can observe a similar pattern to the first row, where the U-Net again converges to the input SNR, and the ResNet results in a lower SNR than the input. In summary, we see that for both networks, it is only when training with data augmentation that we are able to find any clearly visible improvement in SNR over the input for the unseen filter condition.

\subsection{Testing Data}

\begin{table}[t]
\centering

\begin{tabular}{cccccc}
Filter & Experiment & SNR  & $\Delta$SNR  & VGG & $-\Delta$VGG  \tabularnewline   \hline
\multirow{9}{*}{\begin{tabular}[c]{@{}c@{}}Chebyshev1$-6$ \\(seen filter) \end{tabular}}    

& Input       & $27.86$ &         & $46.55$ &         \\
& U-Net       & $30.34$ & $+2.47$ & $41.04$ & $+5.51$ \\
& U-Net DA    & $29.78$ & $+1.91$ & $44.29$ & $+2.26$ \\
& U-Net BN    & $30.90$ & $+3.03$ & $41.52$ & $+5.03$ \\
& U-Net DO    & $30.49$ & $+2.62$ & $41.51$ & $+5.04$ \\
& ResNet      & $34.94$ & $+7.08$ & $39.02$ & $+7.53$ \\
& ResNet DA   & $30.48$ & $+2.62$ & $40.11$ & $+6.43$ \\
& ResNet BN   & $34.37$ & $+6.50$ & $39.41$ & $+7.14$ \\
& ResNet DO   & $\mathbf{35.27}$ & $\mathbf{+7.41}$ & $\mathbf{37.23}$ & $\mathbf{+9.32}$ \\

 \hline
\multirow{9}{*}{\begin{tabular}[c]{@{}c@{}}Butterworth$-6$ \\(unseen filter) \end{tabular}}

& Input       & $27.37$ &         & $47.11$ &         \\
& U-Net       & $28.55$ & $+1.18$ & $41.90$ & $+5.21$ \\
& U-Net DA    & $29.00$ & $+1.63$ & $44.80$ & $+2.31$ \\
& U-Net BN    & $28.77$ & $+1.40$ & $42.06$ & $+5.06$ \\
& U-Net DO    & $28.62$ & $+1.24$ & $42.34$ & $+4.78$ \\
& ResNet      & $21.96$ & $-5.41$ & $47.12$ & $-0.01$ \\
& ResNet DA   & $\mathbf{29.16}$ & $\mathbf{+1.78}$ & $\mathbf{40.52}$ & $\mathbf{+6.59}$ \\
& ResNet BN  
& $23.23$ & $-4.14$ & $46.38$ & $+0.73$ \\
& ResNet DO  
& $22.10$ & $-5.27$ & $46.15$ & $+0.96$ \\

\end{tabular}
\caption{Output signal-to-noise ratio (SNR) and absolute VGG distances (VGG) on the test dataset, and their improvements with respect to the inputs. For SNR, $\Delta$SNR and $-\Delta$VGG higher is better and for VGG lower is better. DA, BN, and DO correspond to data augmentation, batch normalization, and dropout, respectively. The value range of the VGG embeddings and the VGG distances is $0$ to $255$.}
\label{table:results}
\end{table}

As described in Section \ref{sssec:validation}, the validation dataset is small, and the results shown in Figure \ref{fig:all} are calculated and averaged across short excerpts of $8$\,s in duration. In Table \ref{table:results}, we present the performance of our models on the testing data, which now includes the measurement of the SNR and the VGG distance as a perceptual measure, across the entire duration of the test dataset. When testing with the seen filter, the ResNet models without data augmentation outperform all variants of U-Net by at least $4$\,dB, achieving more than a $7$\,dB improvement over the input SNR. The best performing model is ResNet with dropout, improving upon the input SNR by $7.4$\,dB. We also observe that the inclusion of data augmentation reduces performance when evaluated using the seen filter.

When testing with the unseen filter, the two best performing models use our proposed data augmentation method. Here, the ResNet variants without data augmentation produce output SNR levels well below those of the input. The addition of data augmentation improves the performance of both the baseline U-Net and ResNet. Although this improvement is marginal for the U-Net, at $0.45$\,dB, for the ResNet, we observe a much larger improvement of $7.2$\,dB. In testing with the unseen filter, the best performing model is the ResNet with data augmentation, which improves upon the input SNR by $1.8$\,dB.

Considering the VGG distances, the results of the U-Net variants do not change much across different filters. Compared to the seen filter setting, the ResNet variants without data augmentation exhibit worse results with the unseen filter, however, these values are very close to the input value, hence the filter overfitting in terms of the VGG distance is not as severe as the SNR. For the unseen filter setting, while the incorporation of data augmentation worsens the VGG distance by $2.8$ for U-Net, it produces a much larger improvement of $6.6$ for ResNet, making ResNet with data augmentation the best performing model in terms of VGG distance and SNR.

Quantitative results for each test song, along with three audio excerpts can be found at the following link\footnote{\url{https://serkansulun.com/bwe}}.

\subsection{Sample Overfitting}

In Table \ref{table:nooverfit} we present the SNR performance of our baseline models, without any regularization method, on the training and testing data splits separately, and evaluated on all samples in the data splits, across their full duration. To infer whether sample overfitting is occurring (i.e., that the networks are in some sense memorizing the audio content of the training data) we use the seen filter, the 6th order Chebyshev-$1$. For both the baseline U-Net and ResNet, between training and testing data splits, the SNR improvement over the input is very similar suggesting no overfitting to the audio samples themselves.

\begin{table}[ht]
\centering
\begin{tabular}{cccc}

Data split
& Experiment 
& SNR & $\Delta$SNR \\ \hline
\multirow{3}{*}{Training}                               & Input  & 25.99 &                                                                 
\\
                                                    
& U-Net  & 28.34 & +2.35                                                           
\\
                                                    
& ResNet & 33.00 & +7.01                                                            \\ \hline
\multirow{3}{*}{Testing}                                & Input  & 27.86 &                                                                 
\\
                                                    
& U-Net  & 30.34 & +2.48                                                            \\
                                                    
& ResNet & 34.94 & +7.08                                                          
\end{tabular}
\caption{Output signal-to-noise ratio (SNR) of our baseline models, without any regularization on the training and testing data splits separately, and their improvements with respect to the input. The inputs are created using the low-pass filter which was also used during training (the seen filter, $6$th order Chebyshev-$1$).}
\label{table:nooverfit}
\end{table}

\begin{table}[h!]
\centering
\begin{tabular}{cccc}
Model       & \begin{tabular}[c]{@{}c@{}}Number of \\ parameters\end{tabular} & \begin{tabular}[c]{@{}c@{}}Number of \\ MACs\end{tabular} & \begin{tabular}[c]{@{}c@{}}Runtime \\ rate\end{tabular} \\ \hline
U-Net  & 56.4M                                                           & 415.3G                                                   
& 0.14                                                   
\\
ResNet & 55.1M                                                          
& 3609.4G                                                   & 1.06                                                  
\end{tabular}
\caption{Number of parameters, number of multiply-accumulate operations (MACs), and runtimes of our models. The number of MACs roughly corresponds to half of the number of floating-point operations (FLOP). Runtime rate is the time spent in seconds, to process a signal with a length of one second, during testing, i.e., a forward pass where no gradients are calculated.}
\label{table:specs}
\end{table}

\subsection{Visualization of bandwidth extension}

\begin{figure}[t]
\centering
   
\begin{subfigure}{.45\textwidth}
\includegraphics[width=0.99\linewidth]{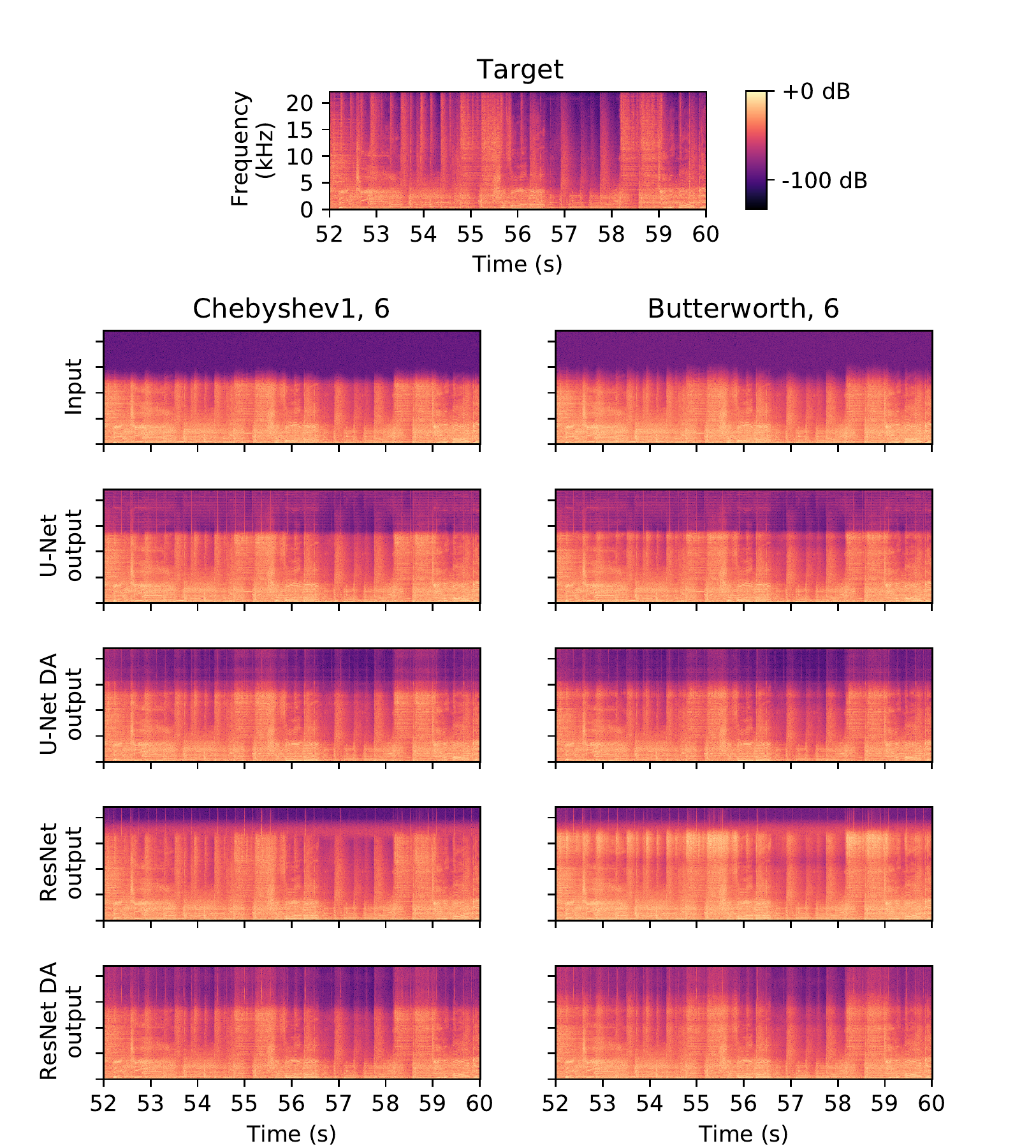}
\caption{Spectrograms of sample audio segments.}
\label{fig:abs_spectrograms}
\end{subfigure}

\begin{subfigure}{.45\textwidth}
\includegraphics[width=0.99\linewidth]{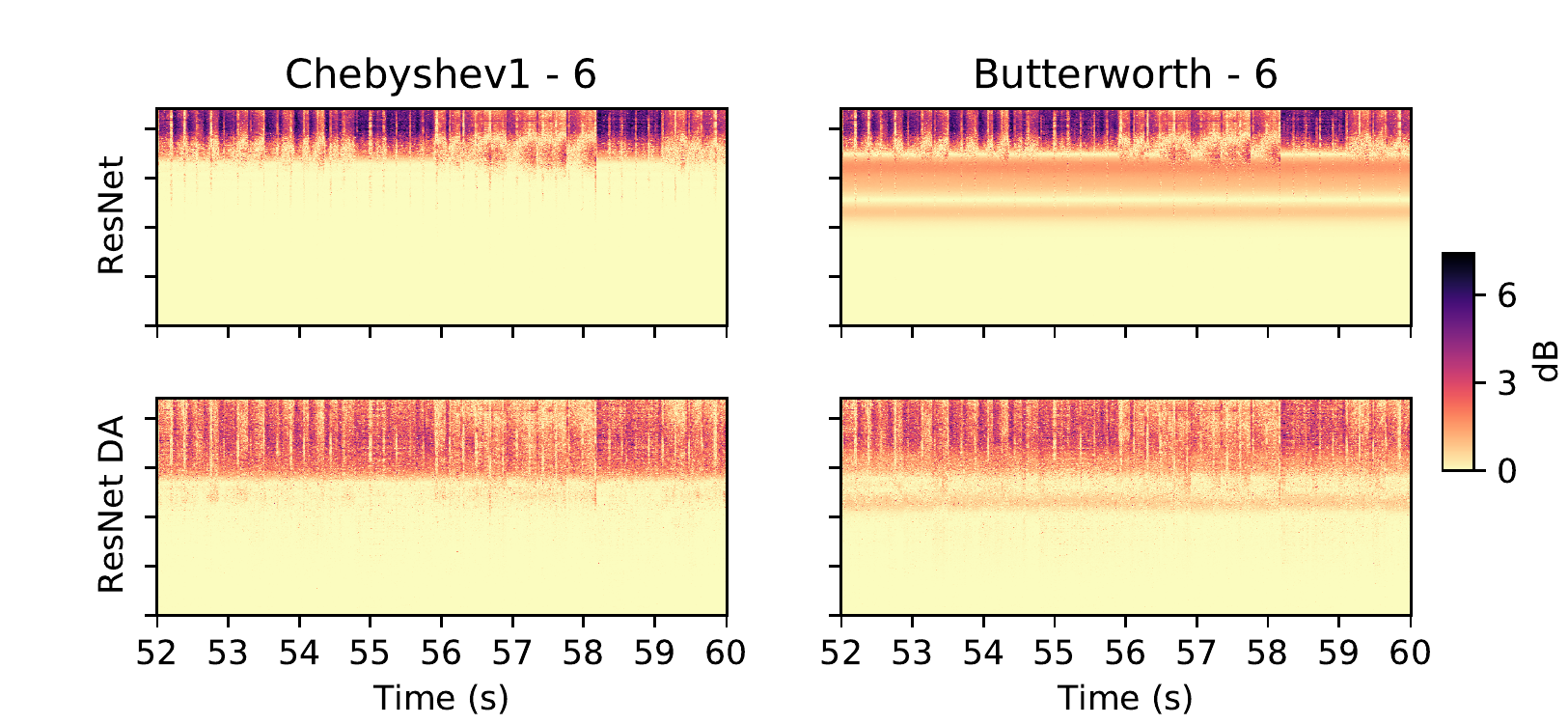}
\caption{Absolute difference with respect to the target spectrogram. The colormap is inverted for better visibility.}
\label{fig:diff_spectrograms}
\end{subfigure}

\caption{Spectrograms and their absolute errors of the sample audio segments. All plots share the axes used in the target (top) plot. Titles per columns denote the type and the order of the filters used. Spectrograms are created using a $1024$-sample Hann window with $50$\% overlap. The audio excerpt is taken from our test set: \textit{DSD100/Mixtures/Test/034 - Secretariat - Over The Top/mixture.wav}}
\label{fig:spectrograms}
\end{figure}

While our proposed method operates entirely in the time-domain, we provide a graphical overview of the outputs of the two networks contrasting the baseline versions with the inclusion of data augmentation for both seen and unseen filters. To this end, we illustrate the spectrograms of one audio excerpt from the test set under each of these conditions in Figure \ref{fig:abs_spectrograms}. The inspection of the figure reveals quite different behavior of the U-Net compared to the ResNet. In general, we can observe more prominent high-frequency information in the output of the ResNet. Of particular note, is the frequency region between approximately $12$-$17$\,kHz for baseline ResNet, and the unseen Butterworth filter, which, contrasting with the target, appears to have ``over-enhanced'' this region. By contrast, once the data augmentation is included, this high-frequency boosting is no longer evident. To emphasize this phenomenon further, in Figure \ref{fig:diff_spectrograms} we display the absolute difference with respect to the target spectrogram, for the baseline ResNet and ResNet with data augmentation. For the unseen Butterworth filter, in the upper half of the spectrogram, the absolute difference of the ResNet with data augmentation is much smoother compared to the baseline ResNet. In this visual representation, we can clearly observe that under all conditions the lower part of the absolute difference spectrogram is essentially unchanged, which reflects the direct additive connection of the input to the output in the network architectures.

\subsection{U-Net vs ResNet: Model Comparison}

As stipulated in Section \ref{sec:models}, we allow both the U-Net and ResNet to have a similar number of parameters. However, we informally observed a distinct difference in training time. In Table \ref{table:specs}, we show several objective properties of these networks, namely the number of parameters, number of multiply-accumulate operations (MACs), and runtimes of our baseline models. Therefore, while both models have roughly the same number of parameters, we see that the U-Net has a much lower runtime and fewer MACs. This is due to its autoencoder-like shape, in which the convolutional layers with more channels are near the bottleneck of the network, where the spatial activation size is the smallest, effectively reducing the number of MACs and the runtime. Looking again at Figure \ref{fig:all}, we can speculate that the ResNet has greater representation power than the U-Net, as shown by its ability to better model multiple known filters than the U-Net, albeit at the cost of slower training and inference.

\section{Conclusions and Future Work}
\label{sec:conclusions}

In this paper, we have raised the issue of filter generalization for deep neural networks applied to musical audio bandwidth extension. Contrary to many problems for which deep learning is used, we do not find any evidence of overfitting to audio samples themselves (i.e. the training data), but rather, we observe a clear trend for state-of-the-art DNNs to overfit to filter shapes. When these DNNs are presented with audio samples that have been pre-processed with low pass filters that do not match the single training filter, then the scope for meaningful extension of the bandwidth is drastically reduced. Furthermore, the use of widely adopted regularization layers such as batch normalization and dropout fall short in alleviating this problem. Looking to the wider context and long-term goal of musical audio bandwidth extension for audio enhancement, we believe that filter overfitting is a critical issue worthy of continued focus.

To address the filter overfitting issue, we have proposed a novel data augmentation approach, which uses multiple filters at the time of training. Our results demonstrate that without data augmentation, filter overfitting increases as training progresses, whereas including data augmentation is a promising step towards achieving filter generalization. While the improvement in generalization for the U-Net is quite small, a more pronounced effect can be observed for the ResNet, which retains high performance across multiple seen training filters. It is particularly noteworthy that the ResNet variants without data augmentation produce very poor results when tested with an unseen filter, with output quality well below that of the input. In this way, the incorporation of data augmentation was the only means to achieve SNR levels that are above the input.

In addition to the primary findings concerning filter generalization, this is, to the best of our knowledge, the first comparison between U-Nets and ResNets in the field of audio processing, and perhaps the first-ever comparison of these approaches given a similar number of parameters. Examining the results of testing with the seen filter, we observe that baseline ResNet outperforms baseline U-Net by a large margin. However, when testing with the unseen filter, baseline ResNet performs the worst.

We argue that the ResNet has more representation power than the U-Net because while the U-Net reduces the spatial activation sizes in its downsampling blocks, the ResNet keeps the spatial activation sizes constant, starting from its input until its output, thus minimizing the loss of information. Even though the networks have the same number of parameters, we can quantify this higher representation power by comparing the number of multiply-accumulate operations. This higher representation power results in the ResNet performing much better in tests with the seen filter, while demonstrating much higher levels of filter overfitting when there is no data augmentation. We show that using the proposed data augmentation method, this powerful network can be successfully regularized, and achieves the best SNR when testing with the unseen filter.

However, if trained without the proposed data augmentation method and tested using an unseen filter, U-Net has less tendency to overfit, making it a more robust network compared to ResNet in this scenario. Furthermore, while we chose to keep the number of parameters within the two models roughly equal, we note that compared to the ResNet, the U-Net is $7.5$ times faster and does nearly $9$ times fewer multiply-accumulate operations (MACs). In this way, the U-Net may be a preferred architecture for real-time streaming applications.

Considering our findings in the broader context of audio enhancement and the potential application to archive recordings, we recognize that low pass filtering alone is by no means sufficient to model the multiple types of signal degradation that can occur. If we wish these trained models to be effective outside of the rather controlled conditions demonstrated here, more work must be undertaken to expand the vocabulary of sound transformations to represent signal degradations including reverberation, wow and flutter, additive noise, and clipping. In this light, the ability of the ResNet with data augmentation to contend with multiple seen filters holds significant promise for a more powerful model to be developed in the future.

A further limitation of our current work is the reliance on SNR and the VGG distance as the indicators of performance. In future work, we consider it of paramount importance to conduct listening experiments to investigate the possible correlations between the subjective evaluations and quantitative perceptual metrics, and to explore models that can improve the perceptual quality such as GANs. Looking beyond the assessment of the perceptual quality of the bandwidth extension, we also seek to investigate listener enjoyment of enhanced archive recordings. Finally, we recognize the potential application of our work on filter generalization to be applied to other types of audio signals, in particular, speech.  

\bibliographystyle{IEEEtran}
\bibliography{references.bib}

\end{document}